# Influence of Ge nanocrystals and radiation defects on C–V characteristics in Si-MOS structures

Shai Levy[*],[a] Issai Shlimak,[a] Avraham Chelly,[b] Zeev Zalevsky,[b] Tiecheng Lu [c]

[a]Department of Physics, Bar-Ilan Institute of Nanotechnology and Advanced Materials, Bar-Ilan University, Ramat-Gan 52900, Israel

[b]School of Engineering, Bar-Ilan Institute of Nanotechnology and Advanced Materials, Bar-Ilan University, Ramat-Gan 52900, Israel

[c] Department of Physics and Key Laboratory for Radiation Physics and Technology of Ministry of Education, Sichuan University, Chengdu 610064, P.R.China

**Abstract**

Metal-Oxide-Semiconductor (MOS) structures containing $^{74}$Ge nanocrystals (NC-Ge) imbedded inside the $SiO_2$ layer were studied for their capacitance characterization. Ge atoms were introduced by implantation of $^{74}Ge^+$ ions with energy of 150 keV into relatively thick (~640 nm) amorphous $SiO_2$ films. The experimental characterization included room temperature measurements of capacitance–voltage (C–V) dependences at high frequencies (100 kHz and 1 MHz).

Four groups of MOS structures have been studied: The 1$^{st}$ - "initial" samples, without Ge atoms (before ion implantation). The 2$^{nd}$ - "implanted" samples, after Ge$^+$ ion implantation but before annealing, with randomly distributed Ge atoms within the struggle layer. The 3$^{rd}$ - samples after formation of Ge nanocrystals by means of annealing at 800$^o$C ("NC-Ge" samples), and the 4$^{th}$ - "final" samples: NC-Ge samples that were subjected by an intensive neutron irradiation in a research nuclear reactor with the integral dose up to $10^{20}$ neutrons/cm$^2$ followed by the annealing of radiation damage.

It is shown that in "initial" samples, the C–V characteristics have a step-like form or "S-shape", which is typical for MOS structures in the case of high frequency. However, in "implanted" and "NC-Ge" samples, C–V characteristics have "U-shape" despite the high frequency operation. In addition, "NC-Ge" samples exhibit a large hysteresis which may indicate charge trapping at the NC-Ge. Combination of the "U-shape" and hysteresis characteristics allows us to suggest a novel 4-digits memory retention unit. "Final" samples indicate destruction of the observed peculiarities of C–V characteristics and recurrence to the C–V curve of "initial" samples.

*Keywords*: Isotopically enriched Ge nanocrystals; neutron-transmutation doping; MOS structures; C–V characteristics;

## 1. Introduction

Capacitance–voltage characteristics (C–V) of Ge and Si nanocrystals embedded in $SiO_2$ films have attracted considerable attention mainly because of their possible application in high-density memory devices [1-4]. In all of the reviewed works, the thickness of $SiO_2$ layer was relatively small (30-50 nm), and semiconductor nanocrystals (NC) were distributed through the insulating matrix between metallic electrode and Si substrate. As a result, the C–V characteristics at high frequencies (0.1–1MHz) had an "S-shape" with large difference between "accumulation" and "depletion" regimes, and exhibit hysteresis caused by recharging of NC when the applied voltage is swept back and forth.

In our work, special samples were fabricated with thick $SiO_2$ layer and narrow distribution of implanted Ge atoms near the upper surface (close to the gate). Our aim was to reduce the influence of Si/$SiO_2$ interface on the C–V curve and focus on the processes occurring in NCs.

[*] Corresponding author. Tel:.+972-3-5317889 ;fax :+972-3-5317749 ; e-mail :shai.levy10@gmail.com.

## 2. Experimental details

p-type silicon (p-Si) crystals with thick (~640 nm) SiO$_2$ layer grown on the <100> oriented surface were used as "initial" samples. Then, $^{74}$Ge$^+$ ions were implanted into the SiO$_2$ matrix with a dose of $1\times10^{17}$ ions/cm$^2$. The ions were accelerated to 150 keV for which the projected range is approximately 100 nm and the struggle (half-width of distribution) is approximately 25-30 nm as predicted by SRIM [5]. These samples were labeled as "implanted" samples. After implantation, some samples were annealed at a temperature of 800$^{\circ}$C. The details of the annealing conditions were published earlier in Ref. [6]. As a result of annealing, randomly distributed Ge atoms in the above layer form nanocrystals (NC) with average diameter of about 4-10 nm. Appearance of NC-Ge was confirmed by the TEM measurements (Fig. 1). These samples are labeled as "NC-Ge". Some of NC-Ge samples were subjected by an intensive neutron irradiation in a research nuclear reactor, followed by annealing of radiation damage. These samples are labeled as "final" samples. Finally, Au contacts were fabricated on top of the SiO$_2$ of all types of samples to create a MOS structure.

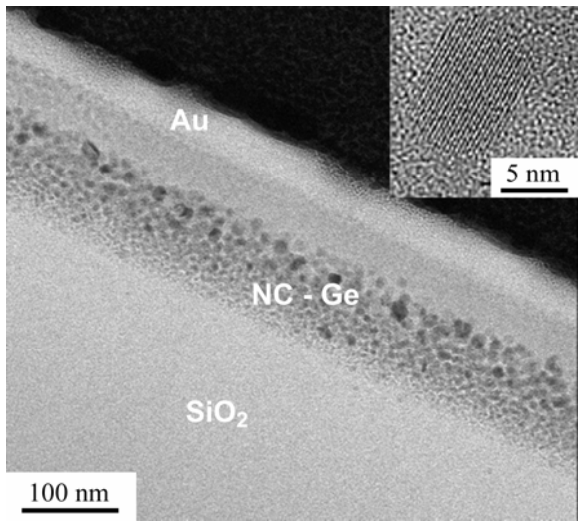

Fig. 1. HR-TEM images of a "final" sample, inset is the HR-TEM image of a Ge nanocrytsal.

## 3. Results and discussion

The C–V characteristic of an "initial" sample exhibits a step-like form or "S-shape" (Fig. 2) which is typical for the p-Si based MOS structures in the case of high frequency measurements. At strong negative voltage, a capacitance corresponds to an "accumulation" regime with maximal value $C_0$ while at transition to positive voltage the capacitance is reduced to a minimal value which corresponds to a "depletion" regime. The low capacitance remains even at strong negative voltages due to the high frequency. In this regime electrons which are minority carriers in p-Si cannot penetrate through the barrier and appear in the "inversion" layer near the interface. Due to large thickness of the oxide layer, relatively high voltages are needed to achieve "accumulation" and "inversion" states, as well as small difference in capacitance is observed between them.

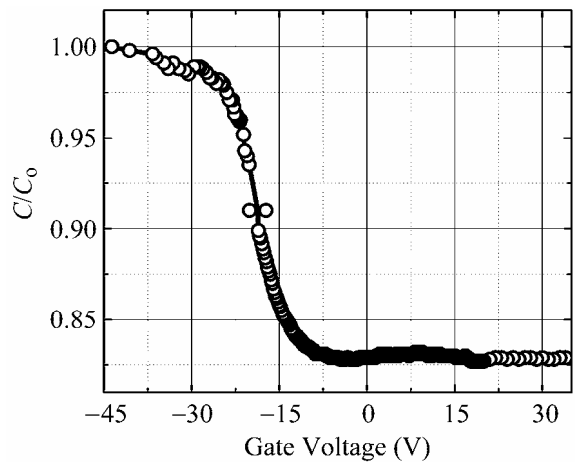

Fig. 2. High frequency (100kHz) C–V curves for "initial" sample.

C–V of "implanted" samples show a significant difference (Fig. 3): the curves have a "U-shape" – at strong positive voltage the capacitance is returned back almost to $C_0$. At first sight, it could be interpreted as manifestation of an "inversion" regime which can be achieved at low frequencies when minority carriers have enough time to appear at Si/SiO$_2$ interface. However, the measurements were done at the unchanged high frequencies (100 kHz and 1 MHz), and besides, implanted Ge atoms (together with radiation defects which accompany the ion implantation) are located too far away from the Si/SiO$_2$ interface to influence in some way the generation-recombination rate of electrons at the interface. Therefore, the "inversion" regime is hardly to be achieved, though one cannot completely exclude this possibility.

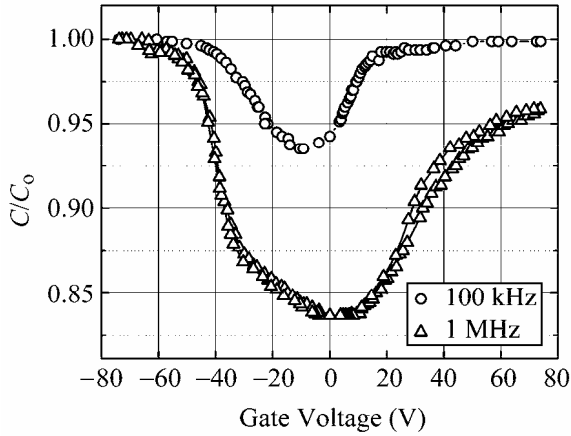

Fig. 3. High frequencies (100 kHz and 1 MHz) C–V curves for "implanted" sample.

a new kind of memory devices with 4-digits memory retention units (four fixed values of voltage at the same value of capacitance). The reason why this effect is observed in our samples is, probably, related to the existence of a relatively narrow layer of NC-Ge in a thick $SiO_2$ matrix and high density of implanted Ge.

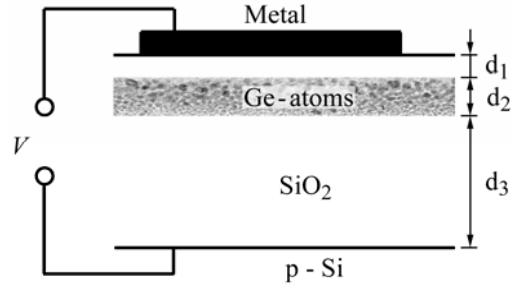

Fig. 4. A schematic general view of a MOS structure after ion implantation where the $SiO_2$ layer is composed from three layers with thickness of $d_1$, $d_2$ and $d_3$.

In our opinion, more realistic is a model based on the polarization of a layer which contains Ge atoms and defects introduced by the implantation. Our MOS structure after ion implantation is composed from three layers with thickness $d_1$, $d_2$ and $d_3$ (Fig. 4) where $d_1$ and $d_3$ correspond to "pure" $SiO_2$ layers, while $d_2$ is the thickness of the "working" layer which is rich of Ge atoms and defects. At zero and weak electric field this layer is neutral and therefore the capacitance is determined by the total width of the structure $C_{min} = \varepsilon/d$ where $\varepsilon$ is the dielectric constant for $SiO_2$ and $d = d_1+d_2+d_3$. In strong field, both negative and positive, the "working" layer can be polarized due to possibility for electrons to tunnel from one atom or defect to another in strongly tilted barriers. If polarization rate is sufficiently high, this layer will not contribute to the total capacitance which now consists of two capacitors in series connection having thickness of $d_1$ and $d_3$: $C_{max} = \varepsilon/(d_1+d_3)$ which is larger than $C_{min}$. It follows that $1 - C_{min}/C_{max} = d_2/d$. One can see in Fig. 3 that $C_{min}/C_{max} \approx 0.9$ which gives $d_2 \approx 0.1d \approx 70$ nm in accordance with the experimental observation (Fig. 1). At the moment this model is more preferable than the first one, but further experiments are needed to make a more conclusive observation.

Fig. 5 shows the C–V of a "NC-Ge" samples when the voltage is swept back and forth between -20V and +20 V. One can see that the C–V curves maintain the "U-shape" but with hysteresis. The existence of hysteresis in MOS with NC-Ge was observed earlier and attributed to the recharging of NC which are located close to the $Si/SiO_2$ interface. This is obtained due to the tunneling of charge carriers from Si substrate. In our samples, NC-Ge are located far away from this interface and charge carriers can tunnel to the trapping centers from the gate. However, in all previous observations, C–V curves were of a step-like shape, while in our work, the hysteresis exhibits a "U-shape" curves which is the first observation to the best of our knowledge. This effect opens the way for suggestion of

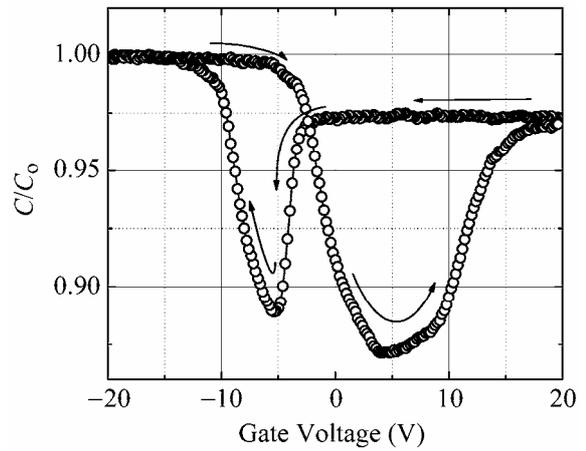

Fig. 5. High frequencies (100 kHz) C–V curves for a "NC-Ge" sample.

Some of the NC-$^{74}$Ge samples were subjected by an intensive neutron irradiation in a research nuclear reactor with the integral dose of $\Phi = 10^{20}$ neutrons/cm$^2$. The neutron flux consisted of slow (thermal) and fast neutrons with energy up to few MeV. Fast neutrons produce radiation damage and can destroy crystallinity in part of the NC-Ge if they collide with a NC or pass in close vicinity (10 nm) from it. The capture of slow (thermal) neutron by isotope $^{74}$Ge leads to transmutation of this isotope into the donor ($^{75}$As) impurity (neutron-transmutation-doping, NTD). The fraction of As atoms $f = N(^{75}As)/N(^{74}Ge) = \sigma(^{74}Ge)\Phi$, where $\sigma(^{74}Ge)$ is the cross-section in cm$^2$ for trapping slow neutron by the given isotope $^{74}$Ge. In our case, $f \sim 10^{-5}$. After irradiation, part of samples were annealed at 800 °C for 30 min to remove the radiation damage ("final" samples). Measurements of these samples show, however, that the "U-shape" of C–V curves before neutron irradiation is not reproduced in spite of the Raman scattering data and TEM images indicating that after annealing, the crystallinity is rebuilt. This can be interpreted as evidence that some peculiarities, which are needed for the observation of "U-shape" characteristics, are destroyed and are not reproduced after annealing of radiation damage. This conclusion needs also further experimental study.

## 4. Conclusions

In this paper we have presented the preliminary experimental mapping as well as the theoretical modeling of C–V characteristics of unique NC-Ge based MOS structure. The behavior of the fabricated device after intensive neutron irradiation was tested.

The proposed device can be used as a new kind of memory devices with 4-digits memory retention units.